\documentclass[10pt,aps,twocolumn,showpacs,prd,preprintnumbers,floatfix,nofootinbib]{revtex4-1}

\usepackage{amsmath}
\usepackage{amsfonts}
\usepackage{graphicx}
\usepackage{amssymb,latexsym}
\usepackage{hyperref}
\usepackage{color}
\usepackage[normalem]{ulem}

\begin{document}


\title{Predictions of just-enough inflation}

\author{Erandy Ramirez}
\email{erandy@tum.de}

\affiliation{Excellence Cluster Universe, 
Technische Universit\"at M\"unchen, Boltzmannstr. 2, 85748 Garching, Germany.}

\author{Dominik J. Schwarz}
\email{dschwarz@physik.uni-bielefeld.de}

\affiliation{Fakult\"at f\"ur Physik, Universit\"at Bielefeld, 
Universit\"atstra\ss e 25, 33615 Bielefeld, Germany}

\begin{abstract}
We find the best-fit cosmological parameters for a scenario of inflation with 
only the sufficient amount of accelerated expansion for the 
$\lambda\phi^4$ potential.
While for the simplest scenario of chaotic inflation all observable primordial 
fluctuations cross the Hubble horizon during the slow-roll epoch, for 
the scenario of just-enough inflation the slow-roll conditions are violated at 
the largest length scales. Performing 
a numerical mode-by-mode integration for the perturbations on the 
largest scales and comparing the predicted anisotropies of the cosmic 
microwave background 
to results from the WMAP 7-yr data analysis, we find the initial conditions
in agreement with current cosmological data. In contrast to the simplest 
chaotic model 
for the quartic potential, the just-enough inflation scenario is not 
ruled out. Although 
this scenario naturally gives rise to a modification of the first 
multipoles, for a quartic potential
it cannot explain the lack of power at the largest angular scales.
\end{abstract}

\date{\today}

\pacs{98.80.Cq}
\preprint{}
\maketitle

\section{introduction}
Early Universe Inflation is considered the physical process behind 
the generation of perturbations that produce the large-scale structures 
observed in the Universe today as well as the mechanism that solves 
the main problems of the standard cosmological scenario. As a proposal 
still lacking a fundamental theory from which it could emerge, there is a 
variety of scenarios which achieve the fundamental aim of inflation 
in different manners.

Current observational results \cite{K10} start to constrain the application 
of the chaotic inflation scenario \cite{linde83} for single-field 
inflation models. A variation of chaotic inflation contemplates 
having only a small amount of accelerated expansion, just enough to solve 
the causality and flatness problems of the hot big bang model. In this scenario,
the initial fast-roll dynamics has observational consequences 
at the larges scales,  
the physics of smaller scales is determined by the subsequent era 
of slow-roll inflation. 
This setup has been considered to explain the lack of power
observed in the lower multipoles of the CMB \cite{cpkl,vsb}. Other 
works have analyzed initial conditions for this scenario in the 
context of a preinflationary radiation-dominated Universe \cite{ws,pk}
and initial conditions for the perturbations different from a Bunch--Davies
vacuum \cite{ddr}. 

In previous works \cite{rs,sr}, we proposed a modification of the chaotic 
inflation scenario also with a limited amount of 
exponential expansion which is in accordance 
with recent observations. Our motivation considered the fact that some
theories of fundamental physics, including the standard model of particle
physics, stop being valid above a certain energy scale. In those cases,
applying chaotic initial conditions would imply a complex potential and 
a decaying inflaton field. By considering an amount of inflation of no 
more than 50 to 60 $e$-foldings of exponential expansion this situation
can be avoided. As a consequence, the Universe starts out in a kinetic-energy-dominated stage with an inflaton potential which is many orders 
of magnitude smaller than the Planck scale to the fourth, but is real 
and well defined. This situation initially causes a violation 
of the slow-roll conditions at the moment of horizon crossing when 
perturbations are evaluated, but the Universe rapidly joins the 
attractor (slow-roll) regime. 

Regardless of this initial violation of slow-roll, we based our previous 
results on the slow-roll expansion. The aim of this work is to formalize the 
implementation of this scenario by integrating the mode equations for
inflationary perturbations in a way consistent with its initial conditions
and find those values that best characterize this scenario in accordance
with current data by means of a Monte Carlo integration.
Here we limit ourselves to the $\lambda\phi^4$ potential only, once the 
the mode integration is carried out correctly, one can consider the 
application of this scenario to arbitrary potentials.

\section{mode integration}

For the purposes of our work, we need to integrate the mode equations for 
the cosmological perturbations for scalar and tensor modes. 
The fact that this scenario of inflation naturally induces a violation 
of scale invariance on observable scales, means that we cannot approximate the 
power spectra and higher-order observables by a power law, as doing 
so would rely on the validity of the slow-roll approximation. 

Our intention is to obtain the initial conditions that predict values for the 
power-spectra and spectral indices which are in accordance with current data.
We therefore need to integrate the mode equations numerically 
and insert this new 
information into the CAMB code \cite{lcl} to supply it with the spectra 
and then into the COSMOMC code \cite{lb}. All relevant information on how 
this is done is provided in the next subsections.

\subsection{Equations of motion : background}\label{back}

As a first approximation, in this work we still assume homogeneity, 
isotropy and flatness from the very beginning of inflation. The equations of 
motion for the background are then 
\begingroup
\everymath{\scriptstyle}
\small
\begin{eqnarray}
\label{ce}
H^2&=& \frac{1}{3M_p^2}\left(\frac{\dot{\phi}^2}{2}+V\right), \\ \nonumber
\ddot{\phi}+3H\dot{\phi}+V'&=& 0;
\end{eqnarray}
\endgroup
a prime denotes a derivative with respect to the scalar field $\phi$, 
a dot means derivative with respect to cosmic time, $H$
is the Hubble rate, whereas $V$ represents the inflaton potential. In this
case, we adopt the notation $V=\frac{\lambda\phi^4}{24}$ and
 $M_p \equiv m_{p}/\sqrt{8\pi} 
\approx 2.4 \times 10^{18}$ GeV is the reduced Planck mass.

We perform all integrations with respect to the number of $e$-foldings $N$, 
which means for the differential equations that 
\begingroup
\everymath{\scriptstyle}
\small
\begin{eqnarray}
\label{de}
\frac{dH}{dN} &=& \frac{V}{M_p^2 H}-3H \\ \nonumber
\frac{d\phi}{dN} &=& -\sqrt{6M_p^2-\frac{2V}{H^2}}
\end{eqnarray}
\endgroup

Our convention is as follows: $\dot{\phi}<0\Rightarrow H'>0$, then
the inflaton roll downs the potential from right to left and
$H \equiv dN/dt$ with $t$ cosmic time, therefore $dN>0$ as $dt>0$.

In what follows, we make use of the horizon flow functions $\epsilon_n$
\cite{stg}, developed as a generalization of the slow-roll parameters
and defined as \footnote{This scheme is completely equivalent to other
schemes in writing the flow equations \cite{klb}.}
\begingroup
\everymath{\scriptstyle}
\small
\begin{eqnarray}
\label{psrpp} 
\epsilon_0 \equiv \frac{H_i}{H}, \quad 
\epsilon_{m+1}\equiv \frac{1}{\epsilon_m}\frac{d\epsilon_m}{dN},\,\,m \ge 0.
\end{eqnarray}
\endgroup
$H_i$ refers to the initial Hubble rate (at $N=0$). In the case of
single-field inflation models, the function $\epsilon_1$ represents
the ratio of the kinetic energy to the total energy density of the
field,
\begingroup
\everymath{\scriptstyle}
\small
\begin{equation}
\label{epsilon}
\epsilon_1 = 3 {\frac{\frac 12 \dot{\phi}^2}{ \frac 12 \dot{\phi}^2 + V}}.
\end{equation}
\endgroup

In the scenario of inflation being considered, the initial value of 
the potential is limited to be much less than the Planck mass such that : 
$V_i \sim M^4\ll M_p^4$, where $M$ represents a scale that bounds 
the validity of the theory under consideration. This means that the evolution
of the system starts in a stage of kinetic energy domination, which later
naturally leads to one of potential energy domination shortly after inflation
starts. Within this picture, it is valid to consider an initial 
condition for the function $\epsilon_{1,i}$ corresponding to a value bigger 
than 1, \cite{linde}, but for a stage of kinetic energy domination,
 an initial condition close to 3 comes out more naturally from 
Eq.~(\ref{epsilon}), as $\dot{\phi}^2_i \sim M_p^4 \gg V_i$. 

\subsection{Equations of motion : perturbations}\label{pert}

In order to calculate the power spectra, we need to integrate the equations 
that describe the evolution for inflationary perturbations. We follow the
notation and conventions of \cite{muk} and write the equations for scalar and
tensor modes in the following manner:
\begingroup
\everymath{\scriptstyle}
\small
\begin{eqnarray}
\label{etau} 
\frac{d^2 u_k^S}{d \tau^2}+\left(k^2
-\frac{1}{z}\frac{d^2 z}{d\tau^2}\right)u_k^S=0
\quad {\rm scalars,}\\ \nonumber
\frac{d^2 u_k^T}{d \tau^2}+\left(k^2
-\frac{1}{a}\frac{d^2 a}{d\tau^2}\right)u_k^T=0
\quad {\rm tensors,}
\end{eqnarray}
\endgroup
where $u$ is defined in \cite{muk} as a gauge-invariant combination of
field and metric perturbations. The comoving wave number $k$ is introduced
to calculate the power spectra through a Fourier expansion. The quantity 
$z$ is defined as $z\equiv a\dot{\phi}/H$, where $a$ is the scale factor
and $\tau$ is conformal time defined as $d\tau=dt/a(t)$.

As mentioned before, we are integrating all quantities with respect to 
the number of $e$-foldings $N$, in consequence, the perturbation equations
are rewritten and integrated as 
\begingroup
\everymath{\scriptstyle}
\small
\begin{eqnarray}
\label{eN}
\frac{d^2 u_k^S}{d N^2}+(1-\epsilon_1)\frac{d u_k^S}{dN}
+\left[\left(\frac{k}{aH}\right)^2
-f_S(\epsilon_1,\epsilon_2,\epsilon_3)\right]u_k^S&=&0,
\\ \nonumber
\frac{d^2 u_k^T}{d N^2}+(1-\epsilon_1)\frac{d u_k^T}{dN}
+\left[\left(\frac{k}{aH}\right)^2-f_T(\epsilon_1)\right]u_k^T&=&0,
\end{eqnarray}
\endgroup
where {\small $f_S(\epsilon_1,\epsilon_2,\epsilon_3)=2-\epsilon_1
+\frac{3}{2}\epsilon_2-\frac{1}{2}\epsilon_1\epsilon_2
+\frac{1}{2}\epsilon_2\epsilon_3+\frac{1}{4}\epsilon_2^2$} and
{\small $f_T(\epsilon_1)=2-\epsilon_1$}, writing 
{\small $\frac{1}{z}\frac{d^2 z}{d\tau^2}$} and {\small 
$\frac{1}{a}\frac{d^2a}{d\tau^2}$} in Eq.~(\ref{etau}) in terms 
of the horizon flow functions (\ref{psrpp}).

When perturbations are generated inside the horizon, in the region where
$k/aH \rightarrow \infty$, the modes approach plane waves of the 
form \cite{muk}
\begingroup
\everymath{\scriptstyle}
\small
\begin{equation}
\label{bd}
u_k(\tau)\rightarrow\frac{1}{\sqrt{2k}}e^{-ik\tau},
\end{equation}
\endgroup
whereas in the opposite limit, when they are outside the horizon
$\frac{k}{aH}\rightarrow 0$, the solution is $u_k\rightarrow z$.

In our case, in order to set the initial conditions for equations 
(\ref{eN}), we need to consider that the system is not in the 
slow-roll regime and we are actually starting the mode integration 
from the very beginning of inflation. At this point, we make another 
approximation which needs to be improved; we set the initial conditions 
for the perturbations assuming the Bunch-Davies vacuum Eq.~(\ref{bd}). 
This clearly needs to be reconsidered as it should not be necessarily the case
that from the very beginning of inflation the part of the Universe on which
inflation starts corresponds to the vacuum state of the field when there is
kinetic energy domination. Our initial conditions are therefore different 
from those adopted in \cite{cpkl} for a similar situation.

We also need to establish a way to determine the value of the first mode
to be integrated, the other modes will be generated from this value. 
We follow the discussion of Ref.~(\cite{muk}), section 11 in which 
a consistent initial condition keeps the terms 
{\small $\left(k^2-\frac{1}{z}\frac{d^2z}{d\tau^2}\right)$}, and 
{\small $\left(k^2-\frac{1}{a}\frac{d^2a}{d\tau^2}\right)$} in Eq.~(\ref{etau})
positive at the beginning of inflation. Therefore the minimum value of
$k$ that we choose is given by the maximum of the initial values between
the terms $\left|\frac{1}{z}\frac{d^2z}{d\tau^2}\right|$ and 
$\left|\frac{1}{a}\frac{d^2a}{d\tau^2}\right|$ which are determined by the
background initial conditions given by this scenario.

The initial conditions we use for the scalar and tensor perturbations for
each mode $k$ in terms of $N$ are

\begin{table}[h!]
\begin{tabular}{c c}
\label{icp}
scalars &
tensors \\

$u_k^S = \frac{1}{\sqrt{2k}}e^{-ik\tau}$, &
$u_k^T = \frac{1}{\sqrt{k}}e^{-ik\tau}$, \\

$\frac{du_k^S}{dN} = -u_k^S\left(\frac{k}{a_iH_i}\right)$, &
$\frac{du_k^T}{dN} = -u_k^T\left(\frac{k}{a_iH_i}\right)$, \\
\end{tabular}
\end{table}

The expressions we use to calculate the power spectra are given by 
\cite{llms}
\begingroup
\everymath{\scriptstyle}
\small
\begin{eqnarray}
\label{ps}
P_R &=& \frac{k^3}{4\pi^2M_p^2}\left|\frac{u_k^S}{z_S}\right|^2, \\ \nonumber
P_T &=& \frac{2k^3}{\pi^2 M_p^2}\left|\frac{u_k^T}{z_T}\right|^2, 
\end{eqnarray}
\endgroup
with $z_S = a\sqrt{\epsilon_1},\,\,\, z_T = a$. Once the first mode $k_{min}$
to be integrated is determined as indicated above, we consider the 
mode $k_{max}=10^5k_{min}$ as the biggest value of interest
for the mode integration. From them we generate 200 equally spaced modes
to perform the integration of equations (\ref{eN}) along with the
background equations (\ref{de}) for each of the modes.

\subsubsection{Integration Method}

We followed the structure of the code in Ref.~\cite{lsv} with the difference
that we do not start the integration at the pivot point $k = aH$ but before
inflation starts and do not Taylor expand any of the quantities being
integrated as we are working with a specific potential. 
The structure of the mode integration as performed in the code was done
in the following manner: given an initial value of $\epsilon_1$ above 1,
one has to find the place where inflation starts and reset the number
of $e$-foldings to 0 to discard the expansion that is noninflationary.
At this moment one determines the value of the first and last modes
that will be integrated with the perturbation equations as described in the
previous paragraph. Then, the background equations are integrated until
the end of inflation to find the values of the potential, Hubble rate,
and quantities that will be used to convert each of the modes, whose 
wave numbers are
given in GeV to units of Mpc$^{-1}$ as needed for the CAMB code. We assume
sudden reheating to establish this correspondence :
\begingroup
\everymath{\scriptstyle}
\small
\begin{equation}
\label{conv}
k_{Mpc^{-1}} = \frac{c}{h}k_{min}\frac{T_{\nu 0}}{T_{end}}\exp^{-\Delta N} 
H_* (1.5\times 10^{38}),
\end{equation}
\endgroup
where $c$ is the speed of light in km/s, $h$ is the Hubble constant 
which we take equal to 70, $k_{min}$ corresponds to the minimum scale in the
CAMB code in units of Mpc$^{-1}$, $T_{\nu}$ is the temperature of neutrinos 
today,
$T_{end}$ the temperature after reheating, $\Delta N = N_{end}-N_*$,
is the number of $e$-foldings between horizon crossing $N_*$ and the end
of inflation $N_{end}$, $H_*$, the value of the Hubble rate at horizon
crossing and the factor $1.5\times 10^{38}$ is used to convert from GeV
to Mpc$^{-1}$. 

In the code, the perturbation equations
and the background equations are integrated beyond horizon crossing
until decaying and growing  modes of scalars and tensors have 
evolved completely and the amplitudes are frozen-in. This happens for tensors 
at $k \approx \frac{1}{14}aH$, a bit later than for scalars,  since the term 
$\frac{1}{a}\frac{da^2}{d\tau^2}$ in Eq.~(\ref{etau}) is bigger than the 
term $\frac{1}{z}\frac{d^2z}{dz^2}$ at the beginning of inflation. 
Consequently we integrate both spectra up to this value. In order 
to optimize the 
computational time, we only 
integrate the perturbation equations until the Universe joins the slow-roll 
regime 
and the amplitudes of both spectra have a difference of less than $10^{-2}$ 
with respect to the corresponding slow-roll expressions. Then we only take 
the later as input into the CAMB code. At the same time, the system of
background
equations is integrated  to find the moment of horizon crossing at $k = aH$ 
and use those values for the conversion of units for the scales. 

\section{monte carlo integration}

One of our main purposes in this work, is to find the best-fit cosmological
parameters for the scenario of inflation mentioned in the Introduction
by considering initial conditions for the background and perturbations
as described at the end of subsection \ref{back} and \ref{pert}. Since this 
scenario produces primordial spectra that differ from the usual power-law
parameterization, we wrote a module that can be used to provide the CAMB code
with the information generated by the mode integration and then be introduced, 
into the COSMOMC code. In our case, the information needed to do the mode 
integration are the initial values of the first horizon flow function 
$\epsilon_{1,i}$, the initial value of the scalar field $\phi_i$, and when 
varied, a range of values for the coefficient $\lambda$ of the potential. 
If this parameter is fixed, then we consider a value of $10^{-12}$ to get 
the amplitude of scalar perturbations correct.

We reused the parameters in the CAMB and COSMOMC codes which are no longer 
needed to parameterize the primordial power spectra and characterize
other inflationary observables since we determine them directly from the 
mode integration \footnote{Thanks to Wessel Valkenburg for his advice
on this point.}.

We use WMAP7 data \cite{K10} only to perform the MCMC analysis. We vary 
the following parameters : the physical densities of baryons, $\Omega_b h^2$, 
and dark matter, $\Omega_{DM} h^2$, and the reionization optical depth $\tau$ 
along with the three parameters mentioned before 
($\epsilon_{1,i}, \phi_i, \lambda$). 
We do not vary the Sunyaev-Zel'dovich template and fix its value to 1. From 
there the 
code also varies the ratio of the sound horizon to the angular diameter 
distance 
$\theta$. The equation of state of dark energy is not varied and is set to 
$w = -1$. The model, as implemented in the COSMOMC package 
delivers the posterior distributions of 13 cosmological parameters, but 
at most 7 of 
them are independent. That is, one extra parameter compared to 
a $\Lambda$ Cold Dark Matter scenario of cosmology implemented 
without varying the Sunyaev-Zel'dovich 
template and without including tensor modes, with the following parameters :
$\Omega_b h^2$, $\Omega_{DM} h^2$, $\theta$, $\tau$, $n_s$ and $A_s$. We 
therefore apply the Akaike information criterion to our results to measure 
the difference in goodness of fit. This is presented below when we discuss 
our results.

\subsection{Initial conditions}

We have mentioned that a natural initial condition for the scenario here 
considered corresponds to the first horizon flow function being close to 3 
which determines the initial value for the Hubble rate through the equation 
\begingroup
\everymath{\scriptstyle}
\small
\begin{equation}
\label{hi}
H_i = \frac{1}{\sqrt{3-\epsilon_{1,i}}}\sqrt{\frac{Vi}{M_p^2}}.
\end{equation}
\endgroup

The initial value of the scalar field determines the initial
value of the potential. From the slow-roll approximation, one can
give an estimate of the interval where the initial value of the scalar
field must vary in order to give the minimum amount of inflation required
to solve the horizon and flatness problems of the standard cosmological
scenario. Here we consider an interval for the field given by 
$\phi_i/M_p = [20,25]$ and let the data decide how to adjust the interval,
after each run.

The code takes into account the fact that, when looking at different regions 
in parameter space for $\epsilon_{1,i}$ and $\phi_i/M_p$, there might be
a numerical issue concerning the way in which the initial value of $k_*$
is chosen; sometimes it could be, that the ratio $k/{aH}$ is too close to 1 
already from the very beginning of inflation. If one is asking the code
to determine the moment of horizon crossing and does the mode integration 
as long as this value is less than 1, the 
numerical accuracy might not be enough to avoid the evaluation being
omitted. The reason is that we are choosing 
$k_{min} = \left(\frac{1}{a}\frac{d^2 a}{d\tau^2}\right)^{1/2}$ which is
bigger than the corresponding term for the scalars as we want to be 
assured that both terms $\left(k^2-\frac{1}{a}\frac{da^2}{d\tau^2}\right)$ 
and $\left(k^2-\frac{1}{z}\frac{d^2z}{dz^2}\right)$ in Eq.~(\ref{etau}) 
are positive. If $\epsilon_{1,i}>1$, $k_{min}$ is chosen at the moment 
when inflation has just started and $\epsilon_1$ has become just one 
time step smaller than 1 and $k_{min}^2\approx a^2H^2$. 
In a situation when the initial 
$\epsilon_1\ll 1$, and the initial $k$ is still chosen in the way
we do for $\epsilon_1 > 1$, then $k_{min}/aH \simeq \sqrt{2}$ which is 
also very close to 1. This problem does not appear in the standard way 
in which the mode integration is done, because the value of 
$k$ is taken within the range reached by observations and the initial 
value of $\epsilon_1$ is considered to be in the slow-roll regime, then 
they are not correlated in the way ours are.

\subsection{Regions explored for the initial conditions}

From Eq.~(\ref{hi}) one can observe that $\epsilon_1 < 3$. 
Therefore an initial value of $\epsilon_1$ could be 
in principle arbitrarily close to 3 from below making the initial Hubble 
rate approach the value of $M_p$ as long as 3 is avoided and the numerical 
accuracy is enough to distinguish them. Apart from this, 
we also explore other possibilities in the Monte Carlo integration. We 
consider values of $\epsilon_{1,i}$ as low as 2 with the consequence that 
one has to reduce the interval on which the initial value of the field can 
vary so that the potential studied in this case can agree with the data. If 
$\epsilon_{1,i} \ll 3$, the initial Hubble rate is some orders of magnitude
below $M_p$ because the potential in this scenario is approximately 8 orders
of magnitude smaller than $M_P^4$.

We also explored the interval [0.001,2.999] for the initial value of
$\epsilon_1$. We switched off the lensing for this run, as it turned out 
that such a broad prior on $\epsilon_i$ is computationally intense, as is 
the lensing itself. Our final conclusions are not based on this run, but it 
provides a nice consistency check.

\section{results}

\begin{table*}
\caption{Summary of main results for Monte Carlo integration showing 
the allowed ranges for the initial values of $\epsilon_1$, the field
and the quartic coupling $\lambda$. For all cases $20\%$ of rows have 
been excluded. Values of $\epsilon_{1,i}$ and $\phi_i/M_P$ outside the 
indicated intervals are disfavored by the data. In all cases the tensor 
modes have been included.}
\begin{tabular}{|c|c|c|c|c|c|c|c|}\hline
\label{iep}
   &
$\epsilon_{1,i}$ &
$\phi_i/M_p$ &
$\lambda$ &
lensing  &
ind. samples &
burn in steps &
R-1\\ \hline
 
1 &
[2.85,2.999] &
[23.55,25.4] &
[$9.28\times 10^{-13}$,$1.26\times 10^{-12}$] &
included &
10 949 &
261 &
0.0017 \\ \hline

 2 &
[2.9,2.999] &
[23.7,25] &
[$9.28\times 10^{-13}$,$1.25\times 10^{-12}$] &
included &
20 058 &
 188  &
0.0023 \\ \hline

3 &
[2.6,2.999] &
[22.74,25] &
[$9.19\times 10^{-13}$,$1.27\times 10^{-12}$] &
included &
7 669  &
424 &
0.0024 \\ \hline

4 &
[0.94,2.44] &
[21.8,25.22] &
[$9.33\times 10^{-13}$,$1.22\times 10^{-12}$] &
not included &
4 307  &
828  &
0.0012 \\ \hline
\end{tabular}
\end{table*}

We explored several regions and cases in parameter space, the most 
representative ones 
are summarized in Table~\ref{iep}. For each case we had to determine 
a new covariance matrix
since we could not use the default covariance matrix supplied with the
COSMOMC program as it would overrun the settings for our initial conditions
\footnote{Thanks to Jussi Valiviita for explaining this issue.}. We
took flat priors on $\epsilon_{1,i}$, $\phi_i/M_p$ and $\lambda$. 

The last column of Table~\ref{iep}, $R-1$, corresponds to the Gelman 
and Rubin R statistic to asses convergence of the chains \cite{gr}. 
These results are for 6 chains of 200,000 samples each. 
We use the features included in the COSMOMC code to propose a good 
covariance matrix but stop updating when the convergence factor R-1 
is less than 0.2 to assure that the chains are strictly Markovian. 
The number of independent samples reported
is obtained by ignoring 20$\%$ of the rows \footnote{Thanks to 
David Parkinson for suggesting this value}.
We found that the number of independent samples is reduced considerably when
the BB mode of the polarization is taken into account. As there are 
no useful observational constraints 
on the BB modes available so far, we restricted our analysis to runs 
with TT, TE and EE spectra only. 

\begin{figure*}[htb!]
\begin{centering}
\includegraphics[width =0.7\linewidth,angle=0]{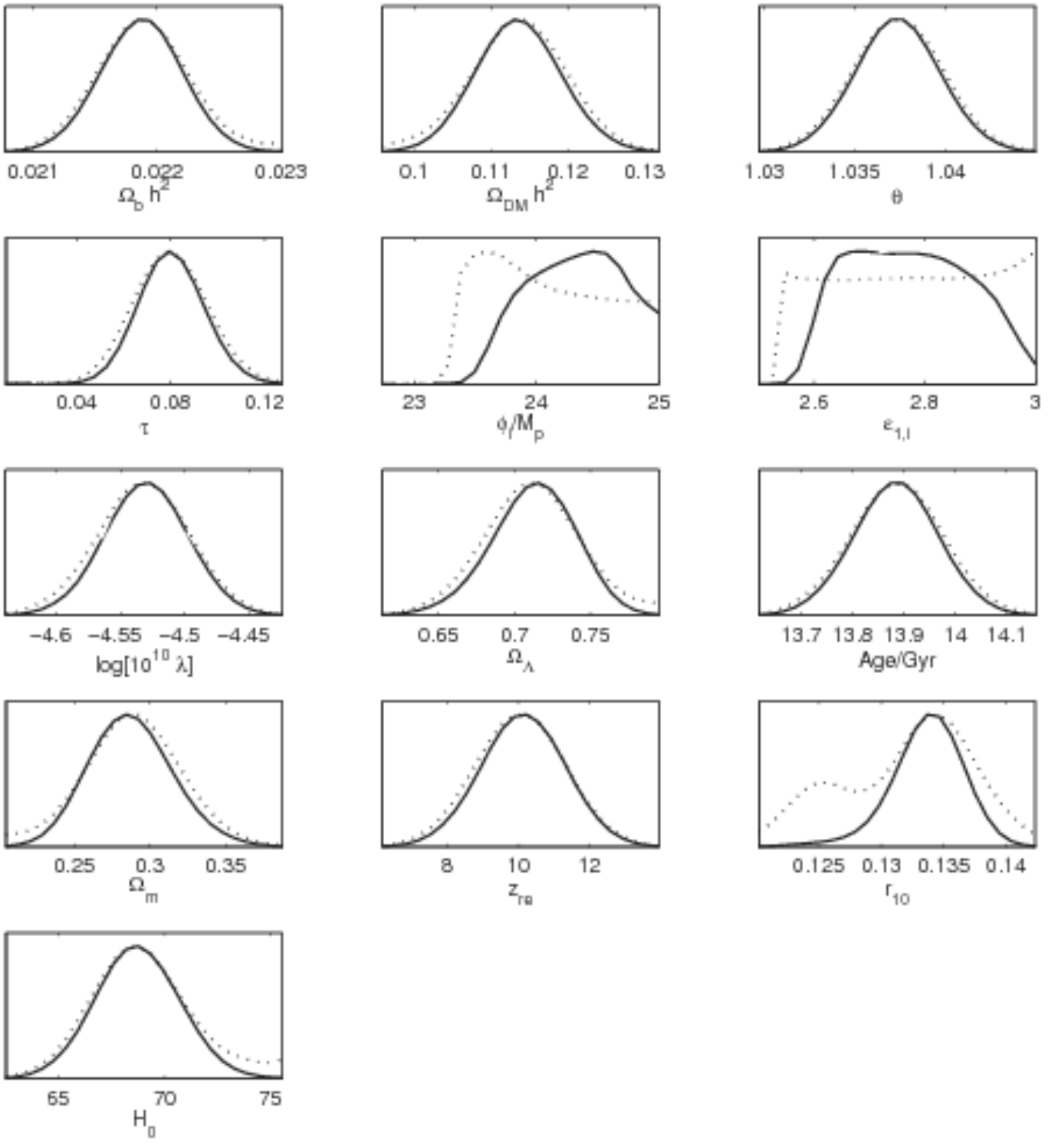}
\caption{Posterior distributions of cosmological parameters from 
case 3 in Table~\ref{iep}. The dotted lines represent mean 
likelihood of samples and solid lines marginalized probabilities.}
\label{f2}
\end{centering}
\end{figure*}

\begin{figure*}[htb!]
\begin{centering}
\includegraphics[width=0.7\linewidth,angle=0]{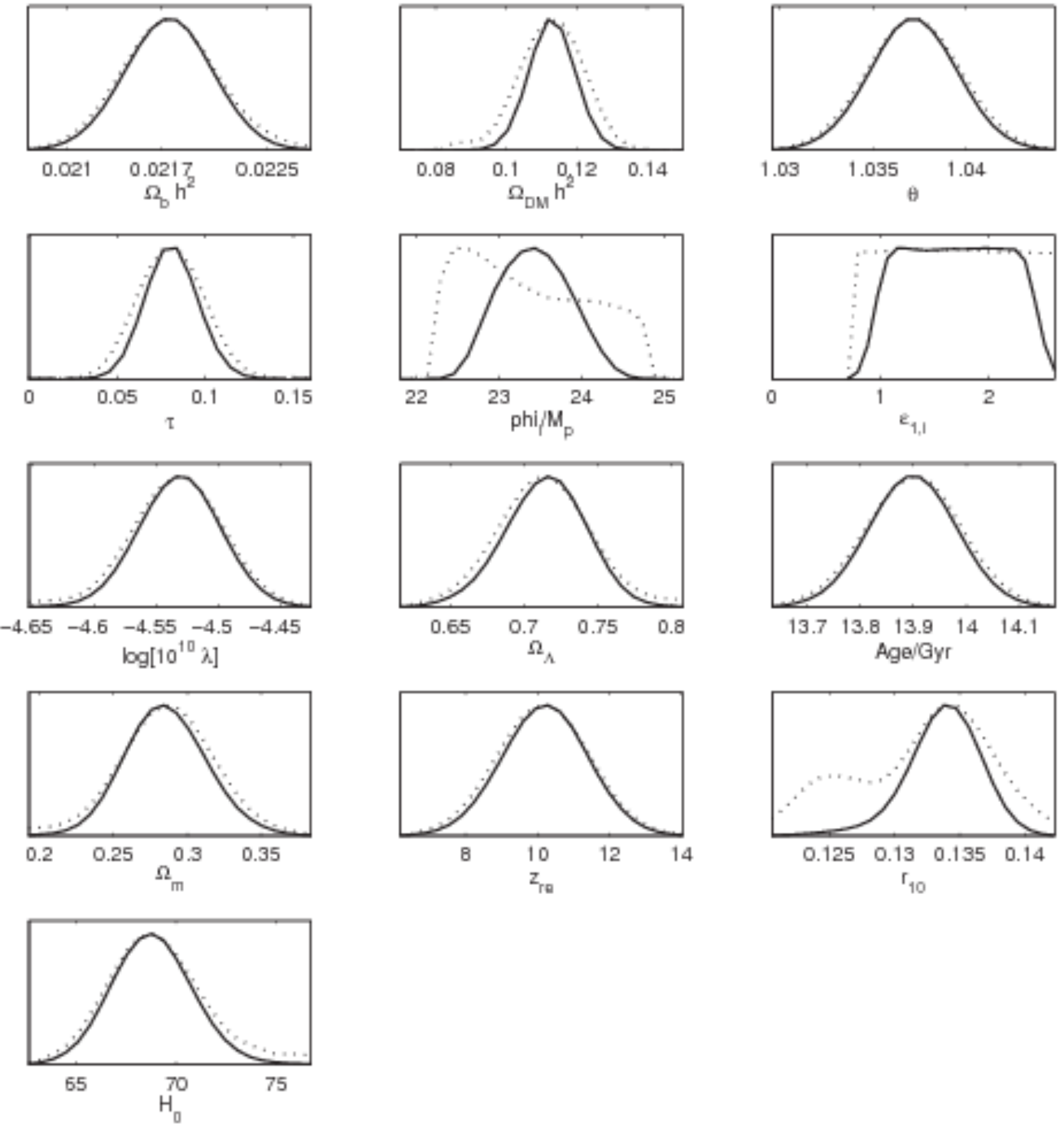}
\caption{Posterior distributions of cosmological parameters from 
case 4 in Table~\ref{iep} .}
\label{f3}
\end{centering}
\end{figure*}

In Table~\ref{iep2}, the mean values and standard 
deviations for the Markov Chains described in Table~\ref{iep} are 
quoted. Their corresponding one-dimensional posterior distributions 
are shown in Figs.~\ref{f2} and 
\ref{f3} only for the last two cases of Table~\ref{iep}.
In all four cases the best fit values of the three `noninflationary' 
parameters $\Omega_b h^2, \Omega_{DM}h^2$ and $\theta$ are in concordance 
with results established by other measurements (supernovae, clusters, etc.). 
The optical depth $\tau$ is clearly measured. The same can be said on the 
self-coupling $\lambda$.  The initial conditions of inflation, encoded 
in $\epsilon_{1,i}$ and $\phi_i/M_P$ are less well constrained. We do find 
good fits to the data (see below), but there is also a degeneracy between 
these two parameters, which results in the discrepancy of the marginalized 
and maximized 1D posteriors in Figs.~\ref{f2} to \ref{f3}, see also 
Fig.~\ref{long_ep} below.

The results for case 1 in Table~\ref{iep} correspond to the lowest possible 
value of $\epsilon_{1,i}$ and lowest and highest values of $\phi_i/M_p$ 
in which the quartic potential is compatible with the data. 
A lower value of $\epsilon_{1,i}$ or wider interval for $\phi_i/M_p$ would
cause all observable modes to cross the horizon during the slow-roll regime 
and thus would result in the exclusion of this potential. Similarly, for 
the cases 2 and 3 a higher value of $\phi_i/M_p$ produces the same effect; 
one starts to recover the slow-roll result for this potential. 

Bigger initial values of the field lead to numerical 
problems when lensing is included. This is not the case when we 
turn off the lensing effect and $\epsilon_{1,i}$ and $\phi_i/M_p$ are 
allowed to vary from slow-roll to 2.999 and [20,26], respectively. One can 
observe in Fig.~(\ref{f3}) that the posterior distribution of  $\phi/M_p$ 
exhibits a maximum. For $\epsilon_{1,i}$, there is a broad region of almost 
equally-likely values. This also repeats in cases 1 and 3, although for 
a narrower interval and different initial conditions.

\begin{table*}
\caption{Summary of main results for Monte Carlo integration. 
Mean values and standard deviations for examples in Table~\ref{iep}.}
\begin{tabular}{|c|c|c|c|c|c|c|c|c|c|c|c|c|c|c|}\hline
\label{iep2}
  &
  &
 $\Omega_b h^2$ &
 $\Omega_{DM}$ &
 $\theta$ &
 $\tau$ &
 $\phi_i/M_p$ &
 $\epsilon_{1,i}$ &
 $\ln(10^{10} A_s)$ &
 $\Omega_{\Lambda}$  &
 Age/Gyr &
 $\Omega_m$ &
 $z_{re}$ &
 $r_{10}$ &
 $H_0$  \\ \hline

1 &
$\mu$ &
0.022 &
0.11 &
1.037 &
0.08 &
24.75 &
2.91 &
-4.53 &
0.71 &
13.89 &
0.29 &
10.14 &
0.13 &
68.75 \\ \hline

 &
$\sigma$ &
0.00032 &
0.0054 &
0.0023 &
0.013 &
0.35 &
0.038 &
0.032 &
0.027 &
0.08 &
0.027 &
1.138 &
0.0028 &
1.92 \\ \hline

 2 &
$\mu$ &
0.022 &
0.11 &
1.038 &
0.08 &
24.65 &
2.93 &
-4.53 &
0.71 &
13.87 &
0.29 &
10.1 &
0.13 &
68.63 \\ \hline

 &
$\sigma$ &
0.00011 &
0.005 &
0.0021 &
0.013 &
0.22 &
0.022 &
0.031 &
0.026 &
0.067 &
0.026 &
1.136 &
0.0031 &
1.86 \\ \hline

3 &
$\mu$ &
0.022 &
0.11 &
1.037 &
0.08 &
24.31 &
2.77 &
-4.53 &
0.71 &
13.89 &
0.29 &
10.13 &
0.13 &
68.75 \\ \hline

 &
$\sigma$ &
0.00032 &
0.0054 &
0.0027 &
0.013 &
0.36 &
0.102 &
0.031 &
0.027 &
0.08 &
0.027 &
1.14 &
0.0029 &
1.92 \\ \hline 

4 &
$\mu$ &
0.022 &
0.11 &
1.037 &
0.08 &
23.43 &
1.69 &
-4.53 &
0.71 &
13.9 &
0.29 &
10.20 &
0.13 &
68.83 \\ \hline
 
 &
$\sigma$ &
0.00032 &
0.0054 &
0.0022 &
0.013 &
0.39 &
0.43 &
0.032 &
0.027 &
0.08 &
0.027 &
1.15 &
0.0028 &
1.94 \\ \hline 
\end{tabular}
\end{table*}

There are some comments concerning the shape of the distributions and
the number of independent samples which are interesting. For 
$\epsilon_{1,i}\in [2.85,2.999]$, $\phi_i/M_p \in [23.55,25.4]$, 
not shown in the table, we obtain very distorted distributions and 
observe that this feature repeats whenever we explore regions of initial
conditions for $\epsilon_1$ below 2.9 and do not include tensor modes.
Particularly, for the interval $\epsilon_{1,i} \in [2.6,2.999]$ without 
tensors and $\lambda$ fixed, also not shown, the posteriors 
are observed to be very distorted, but this is somewhat alleviated by 
introducing the tensor modes and varying $\lambda$, as can be seen from  
Fig.~\ref{f2}. Reducing the initial value of $\epsilon_1$, below 
the favored values in this scenario, causes problems in the simulations.

Regarding the number of independent samples; as mentioned, we have adopted 
for all cases in Table~\ref{iep} to discard $20\%$ of the samples. 
However, in those cases which correspond to $\epsilon_{1,i}$ being almost 3, 
the burn in of the chains occurs earlier. For case 2, discarding 
only $3\%$ of the rows
gives 23 813 independent samples, whereas considering 
$\epsilon_{1,i}\in[2.9,2.999]$ , $\phi_i/M_p\in[23.7,25]$, 
$\lambda\in[9.28\times 10^{-13},1.25\times 10^{-12}]$ with tensors 
 and including the BB mode we find only 11 362 independent samples; 
$\epsilon_{1,i}\in[2.9,2.999]$ , 
$\phi_i/M_p\in[23.66,25]$ and $\lambda = 1\times 10^{-12}$ including only the 
TT mode without tensors discarding $3\%$ of the rows gives 24 328
independent samples. For the other distributions, we did not obtain 
a significant improvement in the
final number of independent samples by repeating the simulation and 
adjusting of the start width 
and standard deviation estimates
in the initial distributions, the numbers shown here were somehow
the best obtained after many attempts. Whether there is a physical argument 
behind this result, that is, whether the number of independent samples 
indicates how physically more likely is a prior distribution, or if it 
indicates the region where the scenario is more favored by the data, we do 
not know.

In Figure~\ref{long_ep} the two-dimensional posterior distributions 
of $\epsilon_{1,i}$ and $\phi_i/M_p$ obtained from the Monte Carlo 
integration for case 4 in Table~\ref{iep} is shown. One can
observe the degree of degeneracy between these two parameters clearly.

\begin{figure}
\begin{centering}
\includegraphics[width=1.1\linewidth,angle=0]{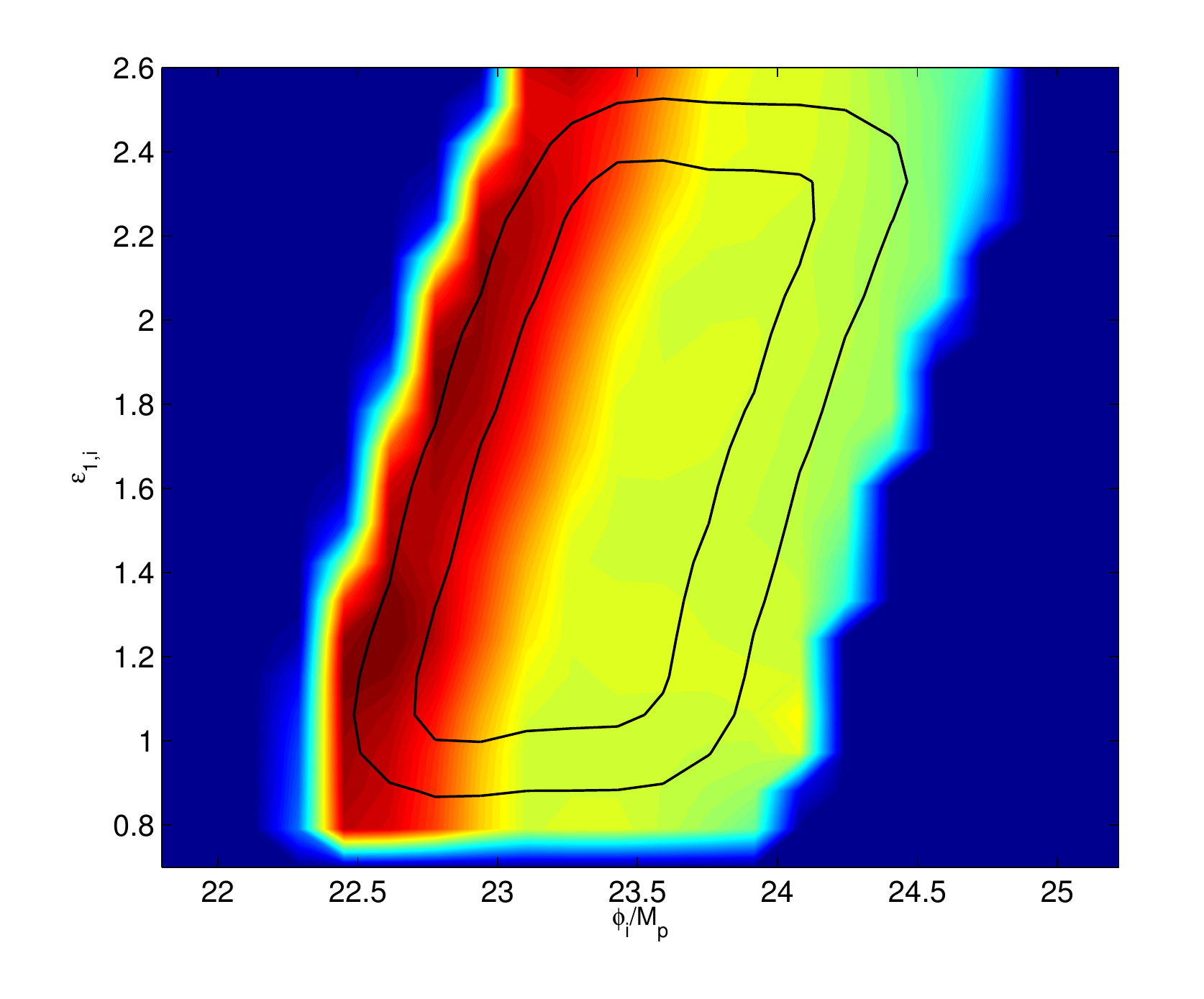}
\caption{(color online). Degeneracy between $\epsilon_{1,i}$ and $\phi_i/M_p$ for case 4
in Table~\ref{iep}. The contours show the 68$\%$ and 95$\%$ confidence  
limits of the marginalized posterior distributions. The shading 
indicates the mean likelihood of samples.}
\label{long_ep}
\end{centering}
\end{figure}

\subsection{Constrained case}

\begin{figure}
\begin{centering}
\includegraphics[width=1.1\linewidth,angle=0]{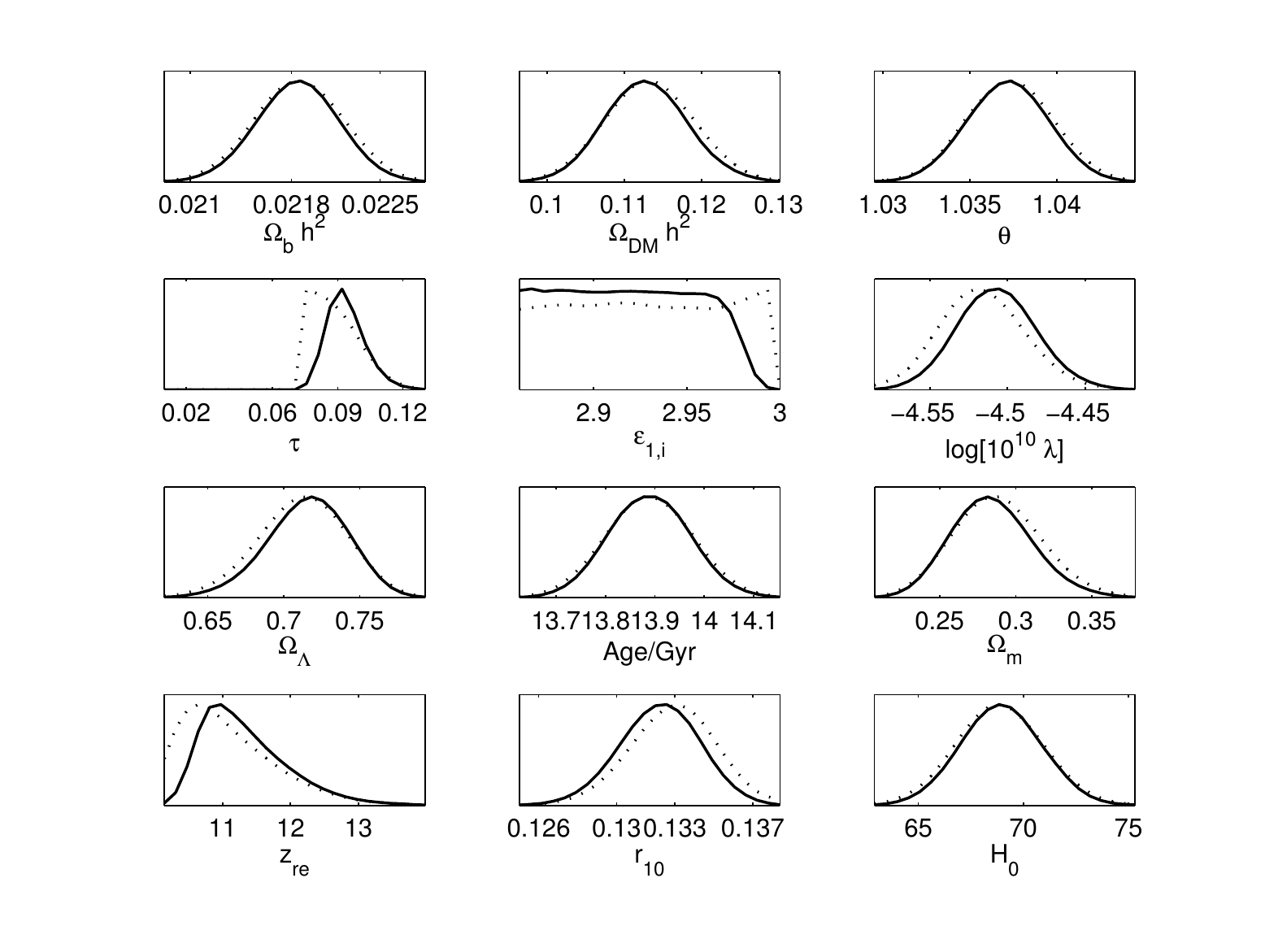}
\caption{Posterior distributions of cosmological parameters from the
 prior $\epsilon_{1,i} \in [2.86,2.98]$, $\phi_i/M_p = 25$, 
$\lambda \in [1.0\times 10^{-12}$,$1.23\times 10^{-12}] $, $C\ell s = 3$ 
 with tensors.}
\label{f10}
\end{centering}
\end{figure}

\begin{figure}
\begin{centering}
\includegraphics[width=1.1\linewidth,angle=0]{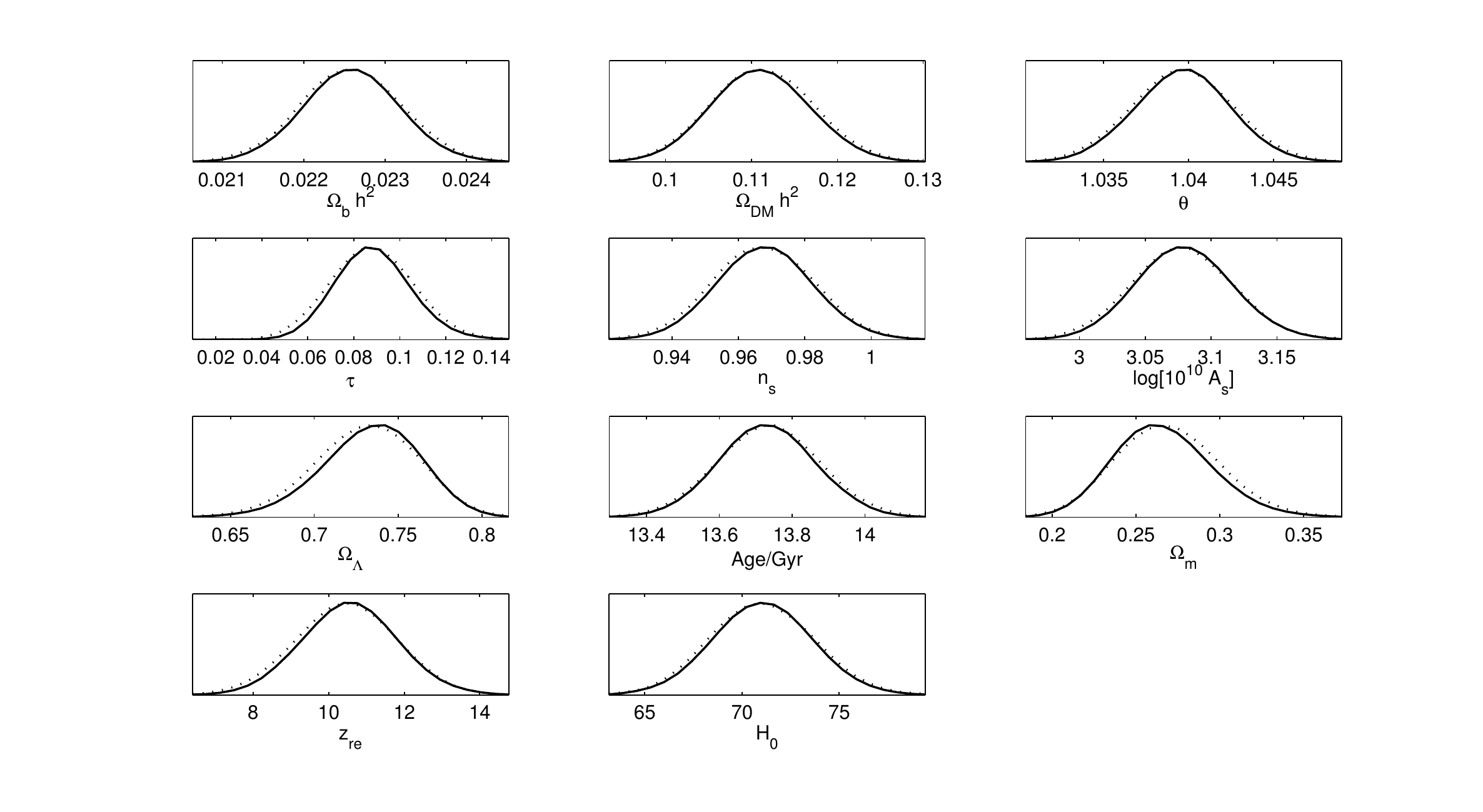}
\caption{Posterior distributions of cosmological parameters for the 
${\rm \Lambda}$CDM model with 6 independent parameters.}
\label{f11}
\end{centering}
\end{figure}

From the results for the distributions of cosmological parameters, 
it is possible to observe that $\epsilon_{1,i}$ and $\phi_i$ are degenerate 
with each other. In order to obtain a better constraints we 
have varied another instance: keeping $\phi_i/M_P$ fixed to a certain value 
and varying $\epsilon_{i,1}$ and $\lambda$, including tensors and lensing
and using the TT, TE and EE angular spectra, which 
is represented by the value of the parameter ($C\ell s=3$).
These results were calculated 
with two chains of 100 000 samples of length each, the best-fit values for 
this case are shown in Fig.~\ref{f10} and the results for a $\Lambda$CDM 
model without tensors and without running is shown in Fig.~\ref{f11}.
 
The posterior distribution produced a flat distribution of $\epsilon_{1,i}$ 
and a central value for 
$\lambda = 1.11\times 10^{-12}\pm 0.03$
with a convergence $R - 1 = 0.0008$, an approximate number of 3921 independent
samples, and burn in of 40 rows. 

We compare our results with a ${\rm\Lambda}$CDM 
model for 6 parameters: $\Omega_b h^2$, $\Omega_{DM} h^2$, $\theta$, $\tau$, 
$n_s$, $A_s$ giving a total of 11 parameters along with $\omega_{\Lambda}$, 
$\omega_m$, $z_{re}$, $H_0$ and age of the Universe. That is, we have  
the same number of primary parameters in the constrained model and one
parameter more in the models reported in Table~\ref{iep}. 
Therefore we apply the Akaike information criterion :
\begingroup
\everymath{\scriptstyle}
\small
\begin{equation}
{\rm AIC} = -2 \ln {\cal L}_{\rm max} + 2 N_{\rm par},
\label{e:akaike}
\end{equation}
\endgroup
in order to compare the goodness of fit of ${\rm\Lambda}$CDM model with 
the four 
models reported in Table~\ref{iep}. In 
Eq.~\ref{e:akaike}, ${\cal L}_{\rm max}$ is the maximum likelihood of 
the model and 
$N_{\rm par}$ is the number of parameters of that model. This is done in 
Table~\ref{akaike} \footnote{Thanks to Zong-Kuan Guo for his explanation.}.

\begin{table}
\caption{Akaike information criterion applied to results of Table~\ref{iep} 
 with respect to a $\Lambda$CDM model of 6 parameters.}
\begin{tabular}{|c|c|c|c|c|c|}\hline
\label{akaike}
Model &
$-\ln {\cal L}_{\rm max}$ &
$N_{\rm par}$ &
AIC &
$\Delta$AIC &
$\chi^2$ \\ \hline

$\Lambda$CDM &
3737.2 &
6 &
7486.5 &
 0 &
7474.4 \\ \hline

1 &
3737.4 &
7 &
7488.8 &
2.3 &
7474.8\\ \hline

2 &
3737.4 &
7 &
7488.8 &
2.4 &
7474.8\\ \hline

3 &
3737.4 &
7 &
7488.8 &
2.3 &
7474.8\\ \hline

\end{tabular}
\end{table}

The main difference between the $\Lambda$CDM model and our results comes
from the number of parameters ($\Delta N_{\rm par} = 1)$. The last 
column in Table~\ref{akaike},  shows the value of a $\chi^2$ goodness 
of fit for these models. 
Compared to the $\Lambda$CDM model, it differs by $0.4$ which assures 
that our scenario 
can fit the data almost as well as $\Lambda$CDM. In terms of the AIC, 
the $\Lambda$CDM with power-law 
spectrum without tensors is slightly favored, but not at a significant level.

As observed from Figs.~\ref{f10}, \ref{f11}, fixing the value of the
field has the consequence of changing the shape of the distributions for the 
reionization redshift $z_{re}$ and the optical depth $\tau$. The initial
value of $\epsilon_1$ however seems to be allowed to include values lower than
2.9 and we have cut the distribution in accordance to the prior imposed, 
although the best-fit value is compatible with the argument that a natural 
$\epsilon_{1,i}$ must be close to 3.

\subsection{Slow-roll  predictions}

In this section, we apply the results for the initial values of 
$\epsilon_{1,i}$ and $\phi_i/M_p$ shown in the last section to
calculate the spectra of anisotropies for the CMB and give the values for the inflationary observables 
calculated at different scales using their slow-roll expressions expanded
at first order. We use the following expression to 
translate the value of the scale $k_*$ which represents horizon
crossing from GeV to Mpc$^{-1}$. This expression was also used
in \cite{rs} for the same purpose and is derived assuming sudden
reheating.
\begingroup
\everymath{\scriptstyle}
\small
\begin{equation}
\label{scale}
k_* = 500 \exp^{-\Delta N_*}H_* \frac{T_{\nu 0}}{T_{reh}}(0.002 {\rm Mpc^{-1}}).
\end{equation}
\endgroup

The assumption of sudden reheating is implicit in the value of $T_{reh}$
calculated from the temperature at the end of inflation. The predictions
for the inflationary observables and the values of the lower multipoles of 
the CMB for each of the cases presented in Table~\ref{iep} are presented 
in Table~\ref{ri} and \ref{cmb} respectively.

\begin{table*}
\caption{Values of inflationary observables 
evaluated at different scales from the result of the Monte Carlo 
integration. The cases are arranged according to the total amount 
of inflation produced by each of them.}
\begin{tabular}{|c|c|c|c|c|c|c|}\hline
\label{ri}
  &
Best-fit values: $\epsilon_{1,i}$, $\phi_i/M_p$, $\lambda$ &
 $N_T$ &
  &
$k_*\,(Mpc^{-1})$,\,\, $N_*$  &
$k_*\,(Mpc^{-1})$,\,\, $N_*$   &
$k_*\,(Mpc^{-1})$,\,\, $N_*$    \\ \hline

 3 &
2.77, \,\, 24.31, \,\, $1.1\times 10^{-12}$ &
64.31 &
 &
0.05, \,\, 59.4 &
$1.03\times 10^{-2}$,\,\,61.0 &
$2.12\times 10^{-3}$,\,\,62.61  \\ \hline

  &
  &
  &
$r$ &
0.27 &
0.26 &
0.28   \\ \hline

  &
  &
  &
$n_s$ &
0.95 &
0.95 &
1.26   \\ \hline

  &
  &
  &
$dn_s/dln k$ &
$-9.0\times 10^{-4}$ &
$-8.42\times 10^{-3}$ &
-0.9   \\ \hline

 4 &
1.69, \,\, 23.43, \,\, $1.1\times 10^{-12}$ &
64.25 &
 &
0.055, \,\, 59.3 &
$1.03\times 10^{-2}$,\,\,61.0 &
$2.15\times 10^{-3}$,\,\,62.6   \\ \hline

  &
  &
  &
$r$ &
0.27 &
0.26 &
0.29   \\ \hline

  &
  &
  &
$n_s$ &
0.95 &
0.95 &
1.3   \\ \hline

  &
  &
  &
$dn_s/dln k$ &
$-9.0\times 10^{-4}$ &
$-9.77\times 10^{-3}$ &
-0.97   \\ \hline

\end{tabular}
\end{table*}

In Table~\ref{ri} $r$, is the tensor to scalar ratio, $n_s$ is the
spectral index of scalar perturbations and ${\rm d} n_s/{\rm d}\ln k$ 
the running
of the spectral index, $N_T$ represents the total amount of inflation
produced by the model. We present the predictions for the observables
on different scales: the first value corresponds to that on which
the Monte Carlo integration was done, $0.05$ Mpc$^{-1}$. The second to the
value that we used in our previous work, $0.01$ Mpc$^{-1}$, the third one is
that of $0.002$ Mpc$^{-1}$, where the analysis of WMAP7 yr is done.
$N_*$ are different points on the integration trajectories with respect
to the number of $e$-foldings where one finds the corresponding value
of $k_*$ by using Eq.~\ref{scale}. In a previous work \cite{rs}, 
we compared the slow-roll trajectories obtained by the WMAP team
to those of this scenario in the $n_s$-$r$ plane. We showed that the modified 
initial conditions lead to an increase of the spectral index at the 
largest scales, in such a way that allows the $\lambda\phi^4$ potential to 
be allowed by current data.

\subsection{CMB Anisotropies}

We intend to see whether the
suppression of power on large scales on the primordial power spectra produced
by this scenario as seen in \cite{rs} could explain the lack of correlation
on large angular scales in the lower multipoles of the CMB, \cite{Spergel},
\cite{Copi}, \cite{Copi2}. There are other works that explore 
the effects of different inflationary setups to explain the lack of power 
at low multipoles of the CMB: using a fitting function and a cut off 
in a fast-roll era \cite{bdvs}, a primordial magnetic field \cite{bh},
anisotropic inflation \cite{gcp}, using a curvaton model \cite{frv} and
isocurvature perturbations, \cite{gh}. In our work, we do not fit the 
primordial power spectra, but allow the numeric solution of the mode 
and background equations to fully determine its shape. We use the initial 
conditions for the background dynamics, determined through the Monte Carlo
integration, to calculate the value of the pivot scale on which perturbations
are evaluated using their slow-roll expressions. The assumption of sudden 
reheating to obtain the value of the pivot scale in $Mpc^{-1}$ gives 
an uncertainty on these numbers. 

There is an important point to mention before presenting the results of this
subsection: We obtained the same values for  $C_{TT}$ for all pivot scales 
$k_*$ considered in Table~\ref{ri}. This is not the case for the standard
power-law parameterization of the primordial spectra. In our case, the
initial conditions for the mode integration are not dependent on the value 
of the pivot scale. We do not use this value  because the scenario
we are investigating is not in the slow-roll regime at horizon crossing
and therefore, setting initial conditions  at horizon crossing, even if
they violate slow-roll do not assure us that we are going to recover
those initial values which characterize this scenario. 

The values of the multipole moments $C_\ell$ do change slightly 
for different 
initial conditions of $\epsilon_1$ and $\phi/M_p$ in Table~\ref{iep}. One
can observe however, that those values are not much lower as compared
to the standard power-law parameterization and initial inflationary
conditions in slow roll. In order to be sure that we were 
giving the values of the mode integration correctly into the CAMB code, 
we used the power-law parameterization of the primordial power spectra 
in our code and the usual slow-roll power-law result
of CAMB without any modifications.  For the same initial value of the 
parameters, we checked that the resulting CMB anisotropies for both instances
at small scales coincided, as was expected. We had to be careful 
not to produce a cut-off in the interpolation since it would affect 
the integration of modes which are deeper subhorizon.

There are some comments to be made in the following about how the
choice of initial conditions and the way in which the primordial
power spectra are included into the CAMB code.

In order to include the new primordial spectra to calculate the
$C_{\ell}$ values, one must interpolate them so that the code can 
find a wider range of values to the whole range of scales reached 
by observations. We used the subroutine SPLINT from
numerical recipes to interpolate. It turns out that, for some initial 
conditions in our scenario, this subroutine produces 
power spectra that are negative. We set this output to 0 as long 
as this happens. For those initial conditions, one has at large scales 
that the value of $C_{TT}$ becomes smaller. Then one can either try to fit 
the quadrupole or octopole separately by trying different initial conditions
as the initial $k$ changes, or a bigger value for the first initial $k$. 
Once this is fixed in the way explained in Sec. \ref{pert}, 
one can choose to use a bigger value for this first scale as it assures that
the terms multiplied by $u_k$ in Eq.~(\ref{etau}) are still positive.
We stress that this is not the value for the pivot scale at horizon crossing 
used for the Monte Carlo integration but the value of the first mode to
be integrated at the beginning of inflation.

If one is exploring parameter space for $\epsilon_{1,i}$ and
$\phi_i/M_p$, one has to be careful not to hit regions in which the amount 
of inflation produced is too small. This can either happen for a fixed 
$\phi_i/M_p$ that gives enough inflation for a certain value of 
$\epsilon_{1,i}$ and as the later goes closer and closer to 3 the amount
of total inflation is reduced, or for a fixed $\epsilon_{1,i}$ and a small
value of the field. Both situations cause the interpolation to produce
again a negative power spectra.

\begin{table}
\caption{Predicted CMB band power $\Delta T_\ell = \ell (\ell+1)C_\ell/2\pi$ 
of quadrupole and octopole for all best-fit models shown in Table~\ref{iep2}.}
\begin{tabular}{|c|c|c|}\hline
\label{cmb}
 &
$\Delta T_2/\mu$K$^2$&
$\Delta T_3/\mu$K$^2$ \\ \hline

1 &
1 228 &
1 139 \\ \hline

 2 &
1 234 &
1 146 \\ \hline

 3 &
1 228 &
1 140 \\ \hline

 4 &
1 228 &
1 140 \\ \hline
\end{tabular}
\end{table}

Having these situations in mind, one can  find either
initial conditions that fit the value of the quadrupole and the octopole
separately, or increase the value of the initial scale to be considered 
for the mode integration so that the final output gives less power
in the lower multipoles, this would be a complementary way to compensate 
for the situation mentioned at the beginning of this subsection. The value
of the pivot scale considered in CAMB does not have any influence in the 
mode integration, but one can consider varying the value of the first scale
to be integrated as long as the equations for the perturbations are well
defined. However, we have not explored further this last situation and its
consequences for the inflationary observables and the Monte Carlo integration.

\begin{figure}
\begin{centering}
\includegraphics[width=1.1\linewidth,angle=0]{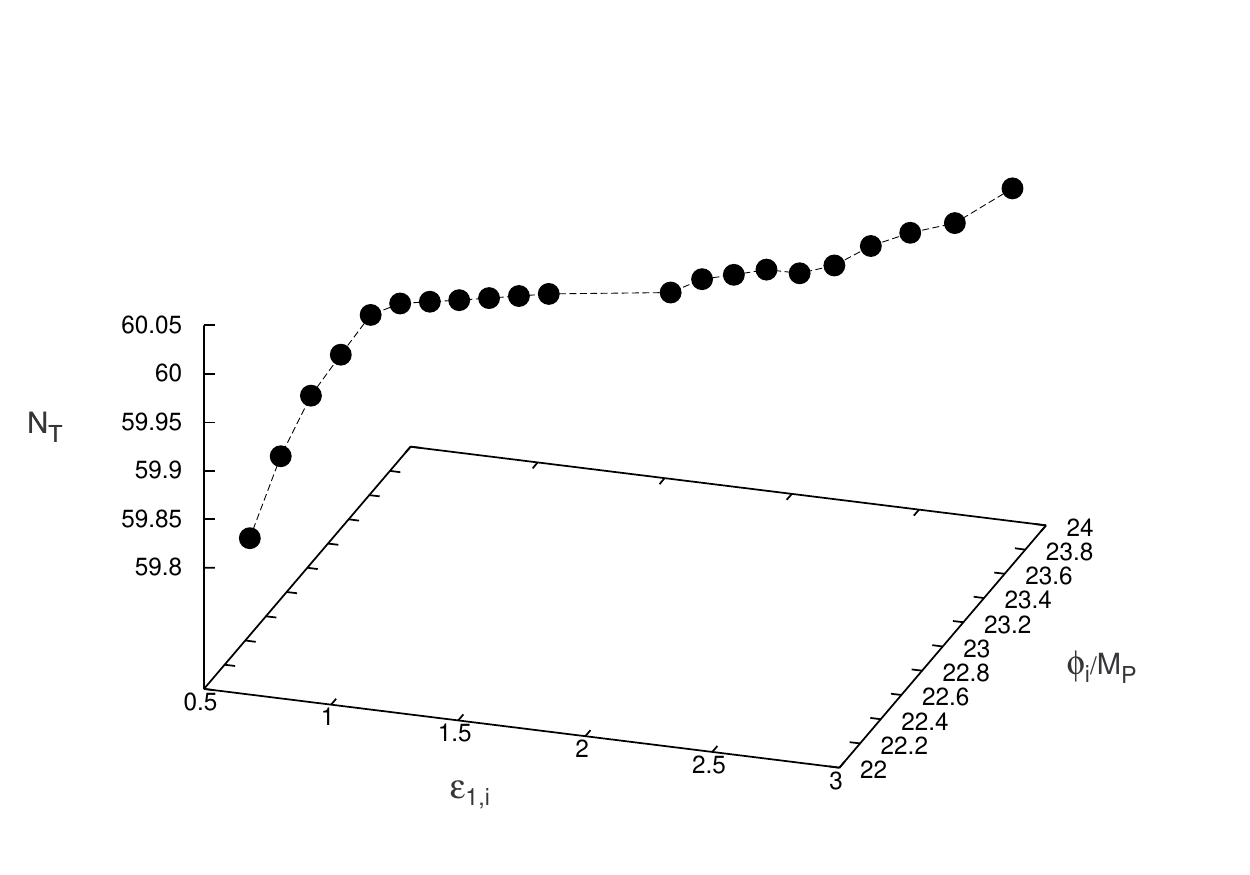}
\caption{Initial conditions in concordance with the observed CMB quadrupole.}
\label{qp}
\end{centering}
\end{figure}

If one tries the first method, the result is that the upper multipoles are 
misplaced, this is different from the
result found in \cite{ws} where they use a $\chi^2$ distribution to fit
the CMB and different initial values for the scale factor considering a 
preinflationary radiation dominated epoch. Those initial conditions that 
enter in the first possibility are listed in Figure~\ref{qp}.

By keeping the value of $\epsilon_{1,i}$ fixed and finding the value of 
$\phi_i/M_p$ that gives the desired result for $\ell = 2$, we found that
as long as $\epsilon_{1,i} > 1$, this happened always at a value of 
$\phi_i/M_p$ that gave the same amount of total inflation. Once 
$\epsilon_{1,i} < 1$ we did not find the same effect. 

By increasing instead the value of the initial scale to be considered 
for the mode integration by a factor of 60, one can reduce the values
of the quadrupole and octopole arriving to values as low as  346.7 
for $\ell = 2$ and 344.7 for $\ell = 3$. The  values are the
lowest ones obtained even if one uses bigger scales. In some of those cases the
interpolation again gives negative values for the output of the primordial 
power spectra and for other values of $k_{min}$, which could be bigger 
or smaller, gives a positive output throughout the whole interpolation process. 
The quadrupole and octopole could no be fit at the same time.

For all the models that fit the quadrupole and octopole by trying different 
initial conditions, there is a range 
of $N_*$ or equivalently, of pivot scales in which the inflationary 
observables $r ,n_s, dn_s/dlnk$ are inside the $1\sigma, 2\sigma, 1\sigma$
intervals respectively, for WMAP7 with running and tensors, but not the
amplitude of scalar perturbations, which turns out to be very small :
$2.11 \times 10^{-9}$ in the best of cases, even for those initial
conditions with $\epsilon_{1,i} < 1$. For the octopole on the contrary, 
the amplitude is too big : $3.31\times 10^{-9}$.

\subsection{Mode Integration results}

Some comments on the results of the mode integration are made in this
subsection. We have mentioned before that for a fixed initial value of
the field, the bigger $\epsilon_{1,i}$ is or closer to 3, the less amount 
of total inflation produced by the model there is. If one keeps fixed 
instead the value of the scalar field, in order to have more inflation 
produced, the value of $\epsilon_{1,i}$ has to be decreased. That is, 
$\epsilon_{1,i}$ and $\phi_i/M_p$ are correlated and the only free 
parameter is the total amount of inflation produced $N_T$. For a fixed 
amount of total inflation,the bigger $\epsilon_{1,i}$ is, the bigger the 
initial value of the field must be in order to compensate.

For the mode integration of perturbations, we fixed the initial
value of the scale factor to 1 at the beginning of inflation. As in
this scenario there is an epoch of noninflationary expansion, we
discarded the number of $e$-foldings during that time and reset it
to 0 at the beginning of inflation. 

\begin{figure}
\begin{centering}
\includegraphics[height=2.3in,width=3.4in,angle=0]{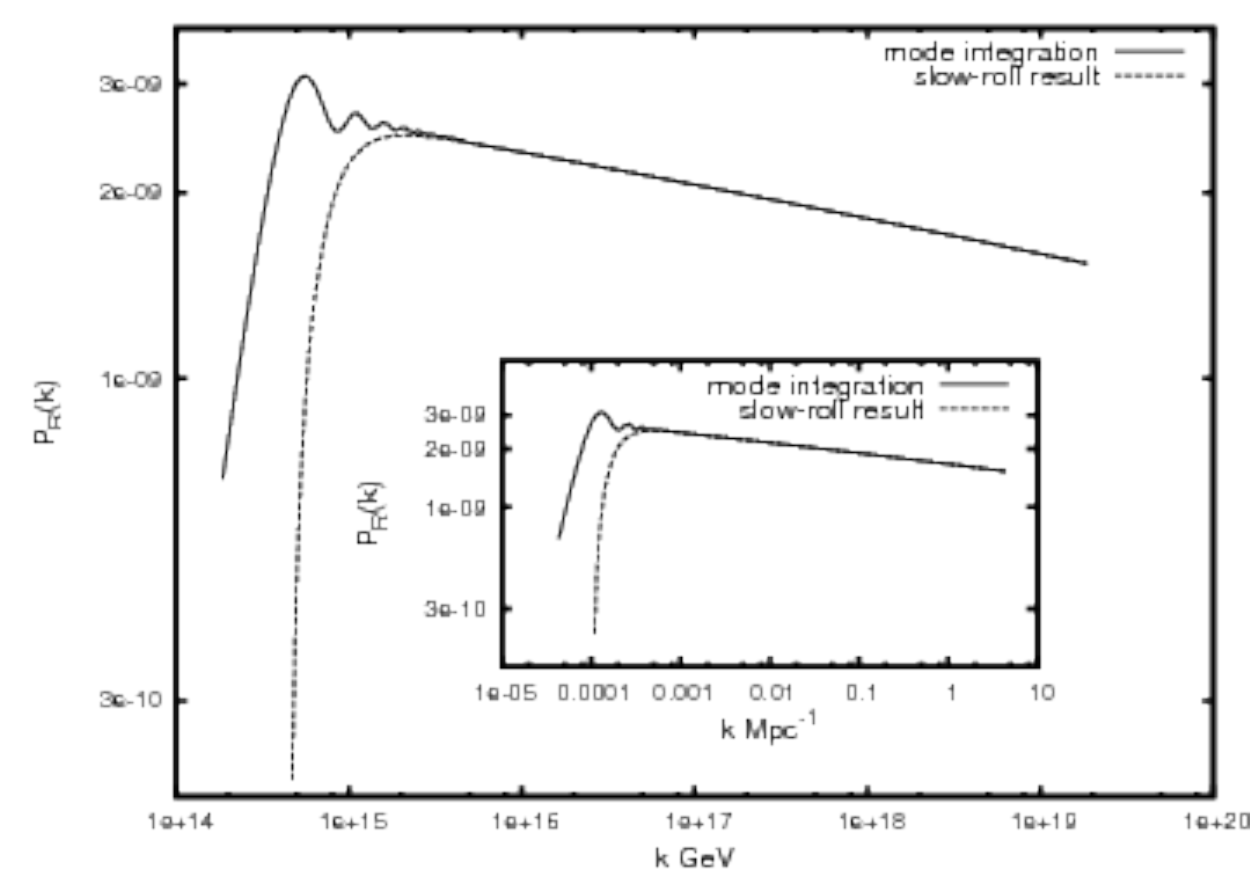}
\caption{Scalar power spectrum for the initial condition 
$\epsilon_{1,i} = 2.93$, $\phi_i/M_p = 24.67$, $\lambda = 1\times 10^{-12}$. 
The main plot shows the scales in 
the x-axis in GeV, the subplot in Mpc$^{-1}$.}
\label{spt_s}
\end{centering}
\end{figure}

\begin{figure}
\begin{centering}
\includegraphics[height=2.3in,width=3.4in,angle=0]{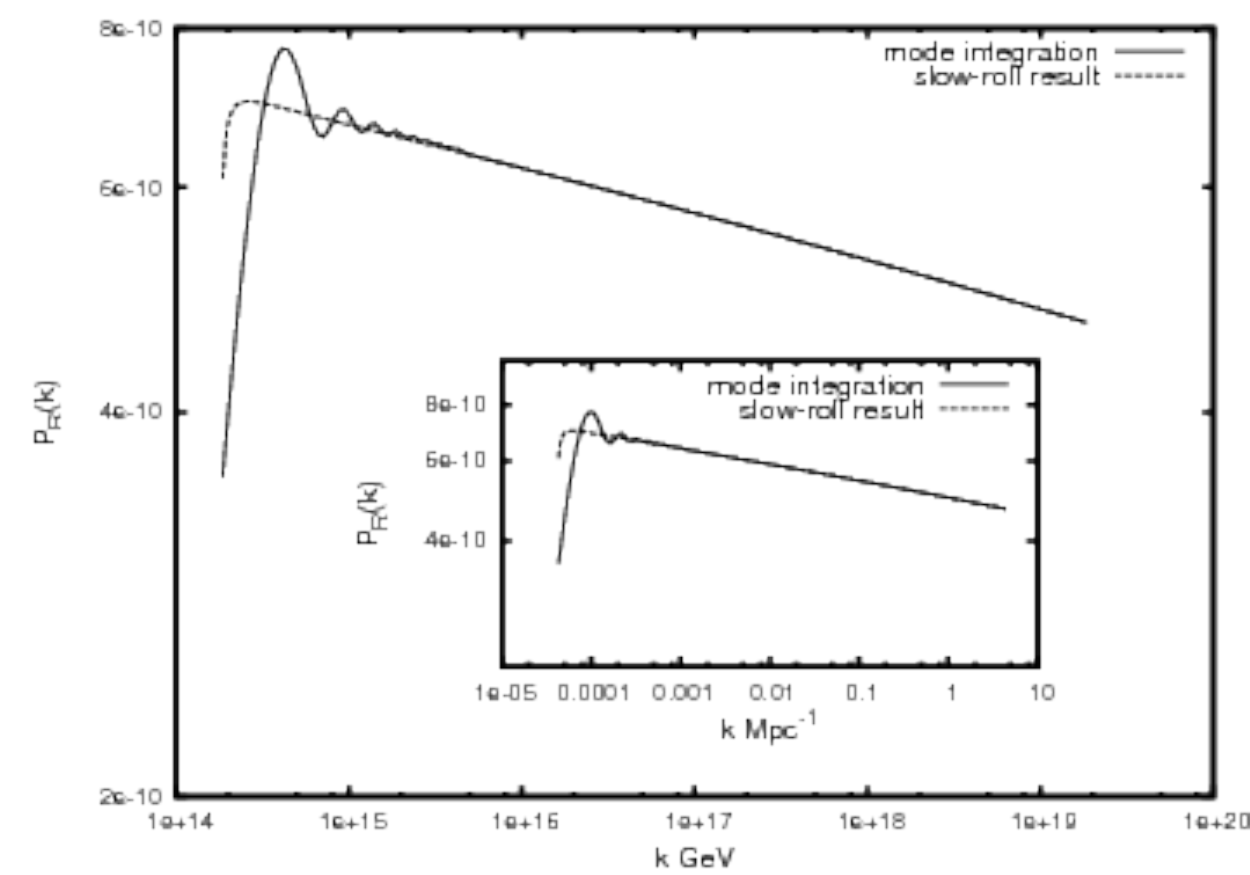}
\caption{Tensor power spectrum for the initial condition 
$\epsilon_{1,i} = 2.93$, $\phi_i/M_p = 24.67$, $\lambda = 1\times 10^{-12}$. 
The main plot shows the scales in 
the x-axis in GeV, the subplot in Mpc$^{-1}$.}
\label{spt_t}
\end{centering}
\end{figure}

In Figs.~\ref{spt_s} and \ref{spt_t} we present examples for the primordial 
scalar and tensor power spectra as obtained from a mode-by-mode integration 
and compared to the corresponding slow-roll predictions.

We can observe oscillations in the amplitude of the scalar and tensor 
perturbations. We have seen before that 
the initial value of the first mode to be integrated is already very close
to horizon crossing due to the way it is chosen for this scenario. This means,
that we are still able to observe the sub horizon oscillations of the 
solutions in Eq.~(\ref{etau}) before the growing mode dominates. In \cite{ws},
it is said that those oscillations disappear by considering an initial 
condition for the scale factor deeper into a previous epoch of radiation 
domination. We are not considering this situation here. However, we have 
checked that if one chooses a bigger initial scale to do the mode integration,
the cutoff and oscillatory behavior observed before disappear completely
and the mode integration and slow-roll spectra are the same for all scales.
That is, one obtains pure power-law spectra.

\section{summary and discussion}

The idea of considering a limited amount of inflationary expansion as a 
modification of the chaotic inflation scenario implies a choice of initial 
conditions that have observable consequences.
In this scenario, the kinetic energy dominates before the onset of inflation 
and consequently violates the slow-roll conditions at horizon crossing 
of observable modes before the system joins the inflationary attractor.

The best-fit power spectrum of scalar perturbations shows a sharp 
cutoff on large scales. This however, does lead to a significant 
suppression  of
the angular CMB power at the largest angular scales. The reason is 
that the cutoff 
in a $\lambda \phi^4$ model with just enough $e$-foldings turns out 
to be at scales that are 
too large to account for a lack of power in the CMB at the largest angles. 
However, one can obtain values very close to those  
of the reported quadrupole and octopole by considering bigger values of 
the initial scale to be integrated, and also by tuning the initial 
conditions to find them separately. This, however, causes higher multipoles 
to have much higher values than observed. However these results are 
only applicable
for the $\lambda\phi^4$ potential.

The results of the mode integration for the tensor modes show that they have
less power, as expected, than the scalars and their power spectrum
also shows a sharp cutoff at large scales.

An interesting point concerning the initial conditions of the
inflaton and $\epsilon_1$ is that the Monte
Carlo integration gives predictions of best-fit values only for those
intervals specified in the priors. It does not give information about
the physical significance of the region explored, unless that information
is contained in the amount of independent samples given by the simulation
for different distributions. We have seen that values
$\epsilon_{1,i} \approx 3$ give bigger numbers of independent samples. 

The Monte Carlo integration confirms in the sense mentioned above, that 
a natural initial condition for this scenario corresponds to the first 
horizon flow function being close to 3. We have already seen that an 
initial epoch of kinetic energy domination naturally leads to this initial 
condition.

We have been able to prove with the results obtained by the Monte Carlo 
integration, that the $\lambda\phi^4$ potential is inside current observational
constraints for single-field models of inflation. 

The implementation of this scenario still lacks the consideration that 
the initial conditions for the perturbations cannot be in the Bunch--Davies
vacuum and that the equations of motion for the background are approximated
by a flat homogeneous Universe. Both should be reconsidered in order to have
consistency with the fact that there is kinetic energy domination at the 
beginning of inflation. However, as a first approach, we could obtain
enough information of the main behavior for the inflationary observables
and the predictions for the CMB anisotropies given by this scenario.
One should be able now to apply it to different potentials and explore more
general situations and their consequences.

\section{acknowledgments}
This research was supported by the DFG cluster of excellence 
"Origin and Structure of the Universe''. We thank the advice 
and help with the numerics of J\"urgen Engels, Rossella Falcone, 
Tommaso Giannantonio, Zong-Kuan Guo, Martin Kilbinger, Julien Lesgourgues, 
Alexey Mints, Aravind Natarajan, David Parkinson, Christophe Ringeval, 
Jussi Valiviita and Wessel Valkenburg. E. R. acknowledges the use 
of the Linux Cluster of the Leibniz-Rechenzentrum 
der Bayerischen Akademie der Wissenschaften and the help 
of Reinhold Bader and Orlando Rivera. We used the CAMB and COSMOMC
packages as well as WMAP 7-yr data from the LAMBDA server.

\end{document}